\documentstyle[12pt]{article}
\begin{document}
\baselineskip=18pt

\centerline {\Large Quantum confinement effects on the ordering of}  
\centerline {\Large the lowest-lying excited states in conjugated chains}

\centerline { \bf Z. Shuai and J.L. Br{\'e}das}
\centerline {Service de Chimie des Matariaux Nouveaux}
\centerline {Centre de Recherche en Electronique et Photonique Mol{\'e}culaires}
\centerline {Universitat de Mons-Hainaut, Place du Parc 20, B-7000 Mons, Belgium}

\centerline {\bf S. K. Pati and S. Ramasesha}
\centerline {Solid State and Structural Chemistry Unit}
\centerline {Indian Institute of Science, Bangalore 560012, India}

\begin{abstract}
	The symmetrized density matrix renormalization group approach is 
applied within the extended Hubbard-Peierls model (with parameters $U/t$,
 $V/t$, and bond alternation $\delta$) to study the ordering of the lowest 
one-photon ($1^{1}B^{-}_u$) and two-photon ($2^{1}A^{+}_g$) states in one-
dimensional conjugated systems with chain lengths, $N$, up to $N=80$ sites. 
Three different types of crossovers are studied, as a function of $U/t$, 
$\delta$, and $N$. The $U$-crossover emphasizes the larger ionic character 
of the $2A_g$ state compared to the lowest triplet excitation. The $\delta$ 
crossover shows strong dependence on both $N$ and $U/t$. The $N$-crossover 
illustrates the more localized nature of the $2A_g$ excitation relative to 
the $1B_u$ excitation at intermediate correlation strengths.
\end{abstract}

    Recently, much attention has focused on the luminescence properties of 
conjugated organic materials because of their potential for application in 
display devices [1]. These studies have underscored the importance of the 
structure of low-lying electronic excited states. Specifically, a major 
parameter is the relative ordering of the lowest dipole allowed singlet 
($1^{1}B^{-}_u$) state and the lowest dipole forbidden singlet ($2^{1}A^{+}_g$)
 state, in the light of Kasha's rule which relates molecular fluorescence 
to the lowest excited singlet state.

    It is well established that correlated electron systems behave differently 
from independent electron systems, especially in the case of excitations. 
Earlier work has shown that the lowest optically forbidden excited state 
$2A_g$ lies below the optically allowed excited state $1B_u$ in polyene 
molecules [2] ( thus preventing any significant luminescence in such 
compounds ), while an independent electron model gives the opposite picture; 
similar results have been found by Periasamy et al. in the case of 
polycrystalline sexithienyl [3] or Lawrence et al. in single crystal 
polydiacetylene [4]. These examples serve as evident manifestation of 
electron correlation in conjugated molecules. 
The influence of electron correlation has also 
been considered as the main origin of lattice dimerization leading to the 
view that conjugated polymers are rather Mott insulators than Peierls 
insulators [5]. In the context of third-order nonlinear optical response 
and photo-induced absorption, the role of higher-lying excited states 
derived from correlated electron models has also been emphasized [6].

     It is also important to stress that the electronic and optical properties
 of conjugated oligomers and polymers differ depending on whether the 
compounds are in the gas phase, in solution, or in the solid state [7]. 
The chemical environment affects the geometric structure as 
well as the electron correlation strength, the latter via dielectric 
screening [8]. Furthermore, the 
characteristics of the conjugation defects present in oligomers 
depend on chain length, which emphasizes the influence of quantum size 
effects [9].

    In view of these features, we believe that these three factors: 
(i) geometric structure, (ii) strength of electron correlation, and 
(iii) quantum confinement, are most relevant for the study of the photo- and 
electro-luminescence response in organic conjugated chains.

     Previous studies of the $1B/2A$ crossover behavior have been carried 
out for short chain systems, so that in the independent electron limit, the 
$2A$ energy is significantly higher than that of $1B$ due to the discreteness 
of the molecular orbital energy spectrum. In this zero $U$ limit, the 
$2A$ state corresponds to single $HOMO$ to $LUMO-1$ (or $HOMO+1$ to $LUMO$) 
excitation while $1B$ is a $HOMO$ to $LUMO$ excitation. According to previous 
results [10], as electron correlation $U$ is turned on, the gap between the 
ground state and the $2A$ state narrows while the gap to the $1B$ state 
increases; the states thus cross at a given Hubbard correlation strength 
$U_c$. This, we refer to as the $U$-crossover. However, for an infinite chain,
 the $2A$ and $1B$ states both occur at the same energy in the H{\''u}ckel 
limit ($U=0$). If the $2A$ and $1B$ states were evolving in a manner 
identical to that in the short chains, these states would never cross with 
increasing $U$. Thus, for a given $U$, their must occur a crossover from 
the short chain behavior to the long chain behavior; this we refer to as 
the $N$-crossover.

     It was noted by Soos, Ramasesha and Galv{\''a}o [11] from exact 
diagonalization studies of short chains that a similar crossover occurred 
with variation of the bond-alternation parameter $\delta$, which we refer 
to as the $\delta$-crossover. The $\delta$-crossover was studied by 
monitoring the optical gap and the lowest singlet-triplet ( spin ) gap; 
the critical $\delta_{c}$ for a given correlation strength was determined 
by the value of $\delta$ at which the optical gap equals twice the 
spin gap. These authors further described the system as behaving band-like 
for $\delta$ values above $\delta_{c}$ and correlated-like for $\delta$ 
values below $\delta_{c}$. However, as was pointed out in Ref. [12], 
increasing bond alternation does not lead to the band picture, because the 
binding energy of the $1B$ exciton increases with increasing $\delta$, 
an obvious indication that electron correlation increases at the same time.

     In this Letter, we present a thorough study that encompasses the three 
kinds of crossovers, namely the $U$, $N$, and $\delta$ crossovers in 
conjugated chains, by employing the 
symmetrized density matrix renormalization group ( SDMRG ) theory. 
The SDMRG approach [13] is currently the most reliable many-body method for 
calculating the low-lying excited 
states with high accuracy for relatively large systems and for 
a wide range of model 
parameters. The model Hamiltonian in this study is the extended 
Hubbard-Peierls Hamiltonian which reads:
\begin{equation}
{\hat H} = \sum \limits_{<ij>,\sigma}{\eta_{i}}
[ - t({\hat a}^{\dag}_{i,\sigma}
{\hat a}_{j,\sigma} + {\hat a}^{\dag}_{j,\sigma} {\hat a}_{i,\sigma})]
 +\sum \limits_{i} \frac{1}{2} U {\hat n}_{i}({\hat n}_{i}-1) \\
+\sum \limits_{<ij>} {\eta_{i}}[V({\hat n}_{i}-1)
({\hat n}_{j}-1)
\end{equation} 
\noindent where $\eta_{i} =[1-\delta(-1)^{i}]$, $\delta$ is the 
dimensionless dimerization parameter, $U$ is the 
on-site Hubbard repulsion ( in units of $t$, the nearest-neighbor hopping 
integral ) and $V$ is the nearest-neighbor charge density-charge density 
interaction. The $\delta$ term serves as a structural parameter in the 
simplest way, if we assume linear electron-lattice coupling in the static 
limit; as has been pointed out before, the $V$-term is crucial to the 
understanding of the optical excitation spectrum, namely the excitonic 
effect [14]. The present model can be regarded as the minimal correlated 
model for conjugated systems. Note that, the meaningful phase corresponds 
to the $BOW$ ( bond-order wave ) regime, namely, $V < U/2$ [15]. By comparing 
experimental data for a series of polyene molecules, we 
find that the parameter set ( $t ,U ,V , \delta$ ) $=$ 
( $2.4$eV, $7.2$eV$=$ $3t$
, $0.4$ $U$, $0.07$) gives the best fit for the 
$1B$, $2A$, and even higher energy $A$ ($mA$) states [16]. We thus set 
$V/U$ $=$ $0.4$ without losing generality. 

     The density matrix renormalization group method is the most accurate 
numerical 
method for determining the ground and low-lying excited states of 
quasi-one-dimensional 
correlated electron systems with short range interactions [17]. 
In the usual DMRG procedure, 
it is difficult to target the $1B_u$ state as there are many states 
that appear between it and the 
ground state, with the number of these states increasing 
with $U$ and chain length $N$. However, 
in a symmetrized DMRG technique that exploits spin parity, $C_2$ 
symmetry, and electron-hole 
symmetry, the $1B_u$ state is the lowest state in the subspace $^eB^{-}_{u}$. 
Incorporating these three 
symmetries thus allows us to determine the $1B^{-}_{u}$ and the 
$2A^{+}_g$ state energies with 
unprecedented accuracy for chains of up to $80$ sites. We choose to 
truncate the space of 
density matrix eigenstates to $100$ ( $m=100$ ) in most cases. For smaller
$U$ and $\delta$, however, we 
choose a larger value of $m$ ($=150$) in order to achieve consistent accuracy.

     We contrast the $U-crossover$ for short ( $N=8$ ) and long ( $N=80$ ) 
chains for fixed 
alternation $\delta=0.07$ in Fig. 1. It is well known that in the strong 
correlation limit, the $2A$ state 
becomes a spin excitation which is gapless in the limit $\delta=0$ and 
this state can be described as 
composed of two triplets. Thus, increase in correlation strength 
should show a decrease in the 
$2A$ energy [18]. However, we note that in the $N=8$ chain, the two-photon 
state energy remains 
nearly constant before decreasing for values of $U/t$ larger than $2.0$. 
In the longer chain, the $2A_g$
energy increases even more rapidly with increasing correlation strength, 
than the $1B_u$ energy. 
This implies a substantial ionic contribution to the $2A_g$ state in long 
chains besides the covalent 
triplet-triplet contribution. This result constitutes the first clear 
illustration of the importance of 
quantum size effects. We find, however, that the critical correlation 
strength, $U_c$, at which the 
crossover occurs is nearly independent of the chain length $N$; in both 
$N=8$ and $N=80$ cases, $U_c$ is around $2.5t$.

For fixed correlation strength ( $U/t$ $=$ $3$ and $4$ ), we present 
the $\delta$-crossover results for 
$N=8$ and $80$ in Fig. 2. We find that the critical $\delta$ value, 
$\delta_{c}$, strongly depends on chain length. 
For $U/t=3$, the $\delta_c$ values are found to be $0.15$ and $0.09$ 
for $N=8$ and $80$, respectively; for $U/t=4$, they are $0.32$ and $0.22$. 
Thus, $\delta_{c}$ has both strong $N$ and $U$ dependence. We also show 
in Fig. 2 
the crossover behavior between the $1B_u$ energy and twice the lowest 
triplet energy, $E_{T}$. This 
crossover occurs at systematically smaller $\delta$ values, again emphasizing 
the larger ionic 
character present in the $2A_g$ state compared to the lowest triplet state.

Most interestingly, we find one more crossover behavior, which is the 
$N$-crossover, in the case of intermediate $U/t$ and medium to large 
$\delta$ values. We observe that the $1B_u$ and $2A_g$ 
states cross over for fixed $U/t$ and $\delta$ as a function of $N$, 
the chain length. The critical lengths 
are actually fairly insensitive to $U$ and $\delta$. 
In Figs. 3a ( $U/t=3$, $\delta=0.12$ ) and 3b ( $U/t=4.0$, $\delta=0.27$ ), 
we find this crossover for $N=14$ and $N=12$, respectively. This is a 
direct theoretical observation 
of quantum confinement induced crossover. It is related to 
the fact that the $2A_g$ excitation is 
more local in character with a shorter characteristic length than the $1B_u$ 
state. Thus, the $1B_u$ 
excitation is stabilized over longer length scales than the $2A_g$ excitation. 
This is seen as a more 
rapid saturation in the $2A_g$ energy compared to the $1B_u$ energy, 
as a function of chain length. 
We note that this crossover can also be seen from Fig. 2 where the 
$\delta_{c}$ 
values show a decrease 
in going from $N=8$ to $N=80$. This behavior can only exist for intermediate 
correlation strength: 
for weak correlation, there does not exist any crossover and $2A_g$ lies 
above the $1B_u$ state for all 
chain lengths as seen from Fig.1; at large values of $U/t$, we are in the 
atomic limit, a crossover 
is not expected and the quantum size effects are largely suppressed. 
It has been widely 
accepted that the conjugated molecules fall in the intermediate 
correlation regime; thus, the 
confinement-induced crossover is realistic.

It is most relevant in this context to stress that, for smaller $\delta$ 
values, even though no 
crossover occurs, the $1B/2A$ gap decreases in long chains. When considering 
the parameters 
adapted to polyenes ( $U/t=3$, $V=0.4U$, $\delta=0.07$ ), both $1B_u$ and 
$2A_g$ state energies decrease with 
$N$, but $2A_g$ state saturates in a way much faster than $1B_u$, because 
the former ( mostly covalent ) 
is more localized than the latter. Consequently, it would be inappropriate 
to extrapolate the 
$1B/2A$ gap in polyacetylene from data obtained on short chains [19]; our 
results imply that the 
actual gap is significantly smaller than the result of such an extrapolation.

To conclude, we have employed the accurate numerical density matrix 
renormalization 
group technique with symmetry adaptation to study the ordering of the 
lowest one-photon and 
two-photon states in conjugated oligomers and polymers within an extended 
Hubbard-Peierls 
model. Three kinds of crossover, namely a $U$-crossover, a $\delta$-crossover, 
and a $N$-crossover, have been demonstrated. The $N$-crossover is related 
to quantum finite size 
effects and crucially depends on the characteristic length of the excitations; 
this characteristic 
length is of the same order as the chain length at intermediate correlation 
strengths and degrees 
of dimerization.

{\bf Acknowledgments}

	This work is partly supported by the Belgian Prime Minister Services 
for Scientific, 
Technical, and Cultural Affairs (Interuniversity Attraction Pole $4/11$ in 
Supramolecular 
Chemistry and Catalysis), FNRS/FRFC (Loterie Nationale), and 
an IBM Academic Joint 
Study. The work in Bangalore is partly supported by the Jawaharlal 
Nehru Center for Advanced Scientific Research.

\newpage

\begin{center}
{\bf References}
\end{center}

\begin{enumerate}
\item J. H. Burroughes, D. D. C. Bradley, A. R. Brown, R. N. Marks, K. Mackay, R. H. Friend, 
P..L. Burn, and A. B. Holmes, Nature 347, 539 (1990); G. Gustafsson, Y. Cao, G. M. Treacy, 
F. Klavetter, N. Colaneri, and A. J. Heeger, Nature 357, 477 (1992).
\item B. S. Hudson and B. E. Kohler, Chem. Phys. Lett. 14, 229 (1972); J. Chem. Phys. 59, 
4984 (1973).
 \item N. Periasamy, R. Danieli, G. Ruani, R. Zamboni, and C. Taliani, Phys. Rev. Lett. 68, 919 
(1992). 
\item B. Lawrence, W. E. Torruellas, M. Cha, M. L. Sundheimer, G. I. Stegeman, J. Meth, S. 
Etemad, and G. Baker, Phys. Rev. Lett. 73, 597 (1994).
\item S. Mazumdar and D. K. Campbell, Phys. Rev. Lett. 55, 2067 (1985); C. Q. Wu, X. Sun, 
and K. Nasu, Phys. Rev. Lett. 63, 2534 (1989).
\item J. M. Leng, S. Jeglinski, X. Wei, R. E. Benner, Z. V. Vardeny, F. Guo, and S. Mazumdar, 
Phys. Rev. Lett. 72, 156 (1994);  F. Guo, M. Chandross, and S. Mazumdar, Phys. Rev. Lett. 
74, 2096 (1995); S. N. Dixit, D. Guo, and S. Mazumdar, Phys. Rev. B 43, 6781 (1991).
\item O. Dippel, V. Brandl, H. Bässler, R. Danieli, R. Zamboni, and C. Taliani, Chem. Phys. 
Lett. 216, 418 (1993).
\item S. R. Marder, C. B. Gorman, F. Meyers, J. W. Perry, G. Bourhill, J. L. Br{\'e}das, 
and B. M. Pierce, Science 265, 632 (1994).
\item J. L. Br{\'e}das and A. J. Heeger, Chem. Phys. Lett. 154, 56 (1989). 
\item a) G. W. Hayden and E. J. Mele, Phys. Rev. B 34, 5484 (1986); b) D. Baeriswyl, D. K. 
Campbell, and S. Mazumdar, in Conjugated Conducting Polymers, edited by H. Kiess, 
Springer-Verlag (Berlin, 1992), and references therein.
\item Z. G. Soos, S. Ramasesha, and D. S. Galv{\"a}o, Phys. Rev. Lett. 71, 1609 (1993).
\item Z. Shuai, S. K. Pati, W. P. Su, J. L. Br{\'e}das, and S. Ramasesha, 
Phys. Rev. B. (1997).
\item S. Ramasesha, S. K. Pati, H. R. Krishnamurthy, 
Z. Shuai, and J. L. Br{\'e}das, Phys. Rev. B, 54, 7598 (1996).
\item D. Guo, et al., Phys. Rev. B 48, 1433 (1993).
\item S. Mazumdar and D. K. Campbell, Phys. Rev. Lett. 55, 2067 (1985); see also Ref. [10b].
\item Z. Shuai, A. B. Saxena, J. L. Br{\'e}das, A. R. Bishop, and B. E. Kohler, to be published.
\item S. R. White, Phys. Rev. Lett. 69, 2863 (1992); Phys. Rev. B 48, 10345 (1993).
\item D. Mukhopadhyay, G. W. Hayden, and Z. G. Soos, Phys. Rev. B 51, 9476 (1995).
\item B. E. Kohler, J. Chem. Phys. 93, 5838 (1990).
\end{enumerate}

\newpage

\begin{center}
{\bf Figure Captions}
\end{center}
\noindent Figure 1. Crossover on $U$ for $\delta=0.07$. 

\noindent Figure 2. Crossover on $\delta$ for (a) $U/t=3$; (b) $U/t=4$. 

\noindent Figure 3. Crossover on $N$ for (a) $U/t=3$ and $\delta=0.12$; 
(b) $U/t=4$ and $\delta=0.27$. 

\end{document}